\documentclass[aps,pre,reprint,superscriptaddress,amsmath,amssymb,longbibliography,floatfix]{revtex4-2}
\usepackage[utf8]{inputenc}
\usepackage{graphicx}
\usepackage{dcolumn}
\usepackage{bm}
\usepackage{textcomp}
\usepackage{url}

\begin{document}

\title{Towards Accurate Prediction of Mutation-Induced Changes in Protein Structure}

\author{Zhuoyi Liu}
\thanks{These authors contributed equally to this work.}
\affiliation{Department of Mechanical Engineering, Yale University, New Haven, Connecticut, 06520, USA}
\affiliation{Integrated Graduate Program in Physical and Engineering Biology, Yale University, New Haven, Connecticut, 06520, USA}

\author{Alex Calabrese}
\thanks{These authors contributed equally to this work.}
\affiliation{Integrated Graduate Program in Physical and Engineering Biology, Yale University, New Haven, Connecticut, 06520, USA}
\affiliation{Department of Molecular Biophysics and Biochemistry, Yale University, New Haven, Connecticut, 06520, USA}

\author{Corey S. O'Hern}
\email{corey.ohern@yale.edu}
\affiliation{Department of Mechanical Engineering, Yale University, New Haven, Connecticut, 06520, USA}
\affiliation{Integrated Graduate Program in Physical and Engineering Biology, Yale University, New Haven, Connecticut, 06520, USA}
\affiliation{Graduate Program in Computational Biology and Biomedical Informatics, Yale University, New Haven, Connecticut, 06520, USA}
\affiliation{Department of Physics, Yale University, New Haven, Connecticut, 06520, USA}
\affiliation{Department of Applied Physics, Yale University, New Haven, Connecticut, 06520, USA}

\begin{abstract}
Proteins can possess numerous mutations relative to their wild-type amino acid sequences with minimal impact to their structure and function. However, in other cases, even a single amino acid mutation relative to the wild-type sequence can lead to a large change in structure or even a disease phenotype. While the accuracy of wild-type protein structure prediction has improved significantly in recent years, it remains difficult to accurately predict the structure of mutant proteins. Here, we characterize the local mutation-induced structural changes in proteins for a dataset of wildtype and the corresponding single-amino acid mutant x-ray crystal structures from the Protein Data Bank (PDB). We find that mutation-induced structural changes in these proteins are localized at the site of the mutation, decaying rapidly with increasing spatial distance from the mutation site. In addition, we evaluate how well AlphaFold3 can recapitulate the observed mutation-induced structural deformations in the x-ray crystal structures. We find that the accuracy of the AlphaFold3 predictions decreases strongly with increasing mutation-induced deformation. In contrast to the results for AlphaFold3, the Pearson correlation between a single physical feature, i.e. the change in solvent accessibility, and the mutation-induced deformation does not depend on the magnitude of the deformation. Our results and analyses provide a framework for further studies aimed at predicting the structural changes in proteins caused by single amino acid mutations.
\end{abstract}


\maketitle


\section{\label{sec:intro}Introduction}

Proteins are heteropolymers composed of sequences of amino acids, which dictate their three-dimensional folded structure and subsequently their biological function. Amino acid sequences contain mutations as a result of errors in DNA replication, RNA transcription, and protein translation~\cite{mordret2019}, such that one in every $\approx 10^3$-$10^4$ amino acids is  mutated~\cite{kramer2007}. However, the protein folding process is robust to mutations with previous studies estimating that $50$-$80\%$ of all possible mutations do not significantly alter the protein's folded structure~\cite{shakhnovich1991}, and only $\approx 35\%$ of the mutations result in functional inactivation~\cite{guo2004}. Yet, there are known disease phenotypes that arise from specific single amino acid mutations. For example, sickle cell anemia is caused by a glutamate-to-valine mutation in hemoglobin~\cite{ramadas2023} that promotes abnormal hemoglobin aggregation, which causes red blood cells to deform into a ``sickle'' shape and become more rigid~\cite{li2017}. Similarly, single amino acid mutations in the human prion protein disrupt the packing of $\alpha$-helices, which promotes protein aggregation and amyloid plaque formation associated with neuro-degeneration~\cite{knaus2001,prabantu2021}.

Many previous studies have investigated whether machine-learning methods can be used to predict the correlation between specific amino acid mutations and human disease states~\cite{choi2012,hecht2015,hopf2017,ng2006,wagih2018} using databases of known links between mutations and disease labels~\cite{landrum2014,landrum2016,landrum2018,harrison2016}. While these studies report that they can accurately predict disease-causing variants with area under the receiver operator characteristic (AUROC) values between $0.8$ and $0.92$~\cite{cheng2023}, predictions of the changes in biophysical properties, such as the binding affinity and thermal stability, of even wild-type proteins remain difficult to predict~\cite{kastritis2010,kastritis2011}. For example, a recent evaluation of nine deep learning models on the ProteinGym deep mutational scanning benchmark found that the best correlations of the predicted changes in biophysical properties upon mutation with the experimental changes had Pearson correlation coefficients ranging from $0.4$ to $0.5$~\cite{tsishyn2025}. These results present a paradox: how can we predict disease outcomes from amino acid mutations, but not changes in the biophysical properties of proteins, when the changes in the biophysical properties are responsible for the disease states. We advocate for a bottom-up approach, where we first understand the structural changes caused by amino acid mutations, then focus on the changes in protein function, and subsequently understand how changes in protein function give rise to disease.

To our knowledge, only a limited number of studies have directly characterized point mutation-induced structural changes at residue-level resolution. Previous work has identified local backbone perturbations following point mutations, but distinguishing mutation-induced changes from the intrinsic thermal fluctuations of proteins remains challenging~\cite{shanthirabalan2018}. It is therefore still unclear how strongly a single-point mutation perturbs neighboring amino acids and how far from the mutation site the wild-type structure is changed. It is also important to determine whether the point mutation-induced changes in structure can be predicted from sequence. Sampling approaches using classical force fields can capture local side-chain rearrangements induced by point mutations, but they have not been employed to model larger backbone conformational changes, where the local root-mean-square deviations in the C$_{\alpha}$ positions ${\rm RMSD} \gtrsim 1$\AA~\cite{smith2008}. More recently, deep learning-based methods have achieved high accuracy in predicting folded protein structures (that are well-represented in the protein Data Bank (PDB)) directly from amino acid sequences~\cite{jumper2021,baek2021,lin2023}, yet studies of their ability to predict mutation-induced structural changes have yielded only modest success. One study found that AlphaFold2 predicted near-wild-type conformations for known destabilizing mutants~\cite{buel2022}. Another study reported a Pearson correlation $\rho \approx 0.4$ between the predicted and experimentally obtained structural changes at the mutation site and that the correlation decayed with distance from the mutation site~\cite{mcbride2023}. Therefore, in this article, we will address several key, open questions concerning the effects of single point mutations on protein structure. 1) How far from the site of point mutations do significant changes in protein structure occur? 2) How accurately can AlphaFold3 predict point mutation-induced structural changes? 3) Can we identify physically interpretable features that can be used to predict mutation-induced structural changes?

To address these questions, we first curated a dataset of x-ray crystal structures from the PDB~\cite{berman2000} of wild-type--mutant sequence pairs in which each wild-type sequence is represented by at least five distinct experimental structures (i.e. at least five x-ray crystal structure `duplicates'). For pairs of x-ray crystal structures, we can quantify the local deformation, $D_i$, for each residue $i$ as the root-mean-square changes in C$_\alpha$ distances between a given residue $i$ and its neighboring residues. We calculate the average {\it mutant} deformation $\bar{D}_i^{\mathrm{mut}}$ by averaging the structural differences over the wild-type duplicate--mutant pairs. We then calculate the normalized mutation-induced deformation $\widetilde{D}_i$ as the ratio of the average mutant deformation to the average duplicate deformation $\bar{D}_i^{\mathrm{dup}}$ obtained by averaging over the deformation of the pairs of duplicate wild-type structures. Quantifying mutation-induced structural changes using $\widetilde{D}_i$ eliminates the effects of the native fluctuations of the protein. We calculate $\widetilde{D}_i (r)$ as a function of $r \equiv |\vec{r}_{m} - \vec{r}_{i}|$, where ${\vec r}_{m}$ is the position of the C$_{\alpha}$ atom at the mutation site $m$. We find that the normalized mutation-induced deformation averaged over all wild-type--mutant sequence pairs, $\langle \widetilde{D}_{i}(r) \rangle$, is largest at the site of the mutation and decays to a plateau value $\gtrsim 1$ for $r > 12$--$15\text{\AA}$. 

We also carried out predictions of the all-atom structure for each mutant sequence using AlphaFold3 and calculated the average predicted mutant deformation, $\bar{D}_i^{\mathrm{mut,pred}}(r)$ by averaging the structural differences between the predicted mutant and experimentally observed wild-type structures. We define the normalized predicted mutation-induced deformation $\widetilde{D}_i^{\mathrm{pred}}(r)$ as the ratio of $\bar{D}_i^{\mathrm{mut,pred}}(r)$ and $\bar{D}_i^{\mathrm{dup}}(r)$. We find that the Pearson correlation coefficient between $\log \widetilde{D}_i^{\mathrm{pred}}(0)$ and $\log \widetilde{D}_i(0)$ is $\rho \approx 0.55$ across all wild-type--mutant sequence pairs, but decreases to $\approx 0.2$ when limited to mutations that give rise to large $\widetilde{D}_i(0)$. We also find that this correlation decreases from $\rho \approx 0.55$ at $r=0$ to $\approx 0.35$ at large $r$. These results indicate that AlphaFold3 is not able to accurately predict large deformations and deformations far from the mutation site. We further identified an important physical feature that is highly correlated with mutation-induced structural changes. We calculate the mean change in the relative solvent accessible surface area upon mutation $\overline{\Delta \mathrm{rSASA}_i}^{\text{ }\mathrm{mut}}$ as the change in $\mathrm{rSASA}$ for residue $i$ averaged over wildtype duplicate--mutant pairs. We then define the normalized change in relative solvent accessibility $\widetilde{\Delta \mathrm{rSASA}}_i$ as the ratio of $\overline{\Delta \mathrm{rSASA}_i}^{\text{ }\mathrm{mut}}$ to $\overline{\Delta \mathrm{rSASA}_i}^{\text{ }\mathrm{dup}}$, where $\overline{\Delta \mathrm{rSASA}_i}^{\text{ }\mathrm{dup}}$ is the change in relative solvent accessibility for residue $i$ averaged over pairs of duplicate wild-type structures. We find that the Pearson correlation $\rho \approx 0.6$ between $\log \widetilde{\Delta \mathrm{rSASA}}_i(0)$ and $\log \widetilde{D}_i(0)$ including all possible wildtype--mutant pairs, and $\rho$ does not decrease with increasing $\widetilde{D}_{i}(0)$. These results show that the simple physical descriptor $\widetilde{\Delta \mathrm{rSASA}}_i$ is strongly correlated with $\widetilde{D}_i(0)$, which can help improve predictions of mutation-induced structural changes. However, $\widetilde{\Delta \mathrm{rSASA}}_i$ cannot be computed without knowing the mutant structure. Since we cannot yet accurately predict the mutant structure from sequence, predicting strongly perturbative mutations remains a major challenge.

\section{Methods}

\subsection{Dataset of Wild-type--Mutant Sequences and their X-ray Crystal Structures}

The generation of the wildtype--mutant dataset consisted of two steps. In the first step, we identified wildtype sequences in the PDB with at least five distinct x-ray crystal structure duplicates. Two sequences were considered the same if their sequences aligned perfectly by pairwise sequence alignment, allowing length differences of up to $10$ amino acids at the N- and C-termini. X-ray crystal structures were excluded if they were part of heteromeric complexes, were bound to nucleic acids, or had crystallographic resolutions larger than $3.5$~\AA. For each wildtype sequence, we then identified sequences with single-point mutations. Sequences containing non-canonical amino acids were excluded, as were structures with missing residues at the mutation site. The final dataset consists of $N_s = 6,967$ wild-type duplicate--mutant sequence pairs, $27,200$ wildtype, and $16,176$ mutant x-ray crystal structures. All x-ray crystal structures in the dataset were processed using PDBFixer to add missing side-chains and hydrogen atoms~\cite{eastman2017}.

\subsection{Local Deformation}

For a given pair of x-ray crystal structures $p$ and $q$, we can quantify the local deformation for each residue $i$ as the root-mean-square deviation in the C$_\alpha$ distances between a given residue $i$ and each neighboring residue $j$\cite{mitchell2016},
\begin{equation}
D_{i}(p,q) = \sqrt{\frac{1}{N(i)}\sum_{j=1}^{N(i)}\left(|\vec{r}_{ij}^{\,p}|-|\vec{r}_{ij}^{\,q}|\right)^2},
\end{equation}
where $\vec{r}_{ij}^{\,p}=\vec{r}_i^{\,p}-\vec{r}_j^{\,p}$, $\vec{r}_i^{\,p}$ is the position of the C$_\alpha$ atom of residue $i$ in structure $p$ and $N(i)$ is the union of the Voronoi neighbors~\cite{rycroft2009} for residue $i$ in structures $p$ and $q$. For residues at the protein surface, the Voronoi cells were truncated at a distance of 2.8~\AA\ away from the surface atoms, approximately the diameter of a water molecule, to focus only on local contributions to deformation.
  
The average duplicate deformation of residue $i$ for a given wild-type sequence is
\begin{equation}
\bar{D}_{i}^{\mathrm{dup}} = \frac{2}{N_{d}(N_{d}-1)}
\sum_{p=1}^{N_d-1}\sum_{q=p+1}^{N_d} D_{i}(p,q),
\end{equation}
where $N_d$ denotes the total number of duplicate structures for each wild-type sequence and the summation runs over all unique pairs of duplicate structures for a given wild-type sequence. Similarly, the average mutant deformation of residue $i$ for a given mutant sequence is 
\begin{equation}
\bar{D}_{i}^{\mathrm{mut}} = \frac{1}{N_{m}N_{d}}
\sum_{p=1}^{N_{m}}\sum_{q=1}^{N_{d}} D_{i}(p,q),
\end{equation}
where $N_m$ is the number of mutant structures for a given wild-type--mutant sequence pair. 
We then define the normalized mutation-induced deformation of residue $i$ as
\begin{equation}
\widetilde{D}_{i} = \frac{\bar{D}_{i}^{\mathrm{mut}}}{\bar{D}_{i}^{\mathrm{dup}}},
\end{equation}
which quantifies the average mutant deformation (for a given wild-type--mutant sequence pair) relative to the intrinsic structural fluctuations among wild-type duplicate structures. Note that $\bar{D}_{i}^{\mathrm{dup}}$ possesses a peak at the mutation site indicating that researchers typically place single point mutations in flexible, solvent-exposed regions for the purpose of crystallizing mutant protein structures. (See Appendices A and B.)

To characterize the spatial extent of mutation-induced structural deformation, we define $r \equiv |\vec{r}_m-\vec{r}_i|$ as the C$_\alpha$ distance between residue $i$ and the mutation site $m$. We can therefore calculate $\bar{D}_i^{\mathrm{dup}}(r)$, $\bar{D}_i^{\mathrm{mut}}(r)$, and $\widetilde{D}_i(r)$ as a function of the distance between residue $i$ and the mutation site. For example, $\widetilde{D}_i(0)$ represents the normalized mutation-induced deformation at the mutation site.

We further defined the $Z$-score of the mutation-induced deformation of residue $i$ as
\begin{equation}
Z_{i} = \frac{\bar{D}_{i}^{\mathrm{mut}} - \bar{D}_{i}^{\mathrm{dup}}}
{\sigma_i^{\mathrm{dup}}},
\end{equation}
where
\begin{equation}
\sigma_i^{\mathrm{dup}} = 
\sqrt{
\frac{2}{N_d(N_d-1)}
\sum_{p=1}^{N_d-1}\sum_{q=p+1}^{N_d}
(D_i(p,q)-\bar{D}_i^{\mathrm{dup}})^2
},
\end{equation}
is the standard deviation of the local deformation over all unique pairs of wild-type duplicate structures for residue $i$. The $Z$-score quantifies the magnitude of the mutation-induced deformation relative to the variability of the intrinsic structural fluctuations among wild-type duplicate structures.

\subsection{AlphaFold3 Prediction of Mutant Structures}

We predicted the all-atom structures of $6,967$ mutant sequences using AlphaFold3 version 3.0.1~\cite{abramson2024}. For each mutant sequence, all-atom structures were predicted using up to $N_a = 20$ random seeds, with each seed generating five candidate structures. For each seed, we selected the candidate structure with the highest ranking score and used these $N_a$ structures as the predicted mutant structures for subsequent analysis. We verified that increasing $N_a$ does not affect the accuracy of the predicted deformation of the mutant structures. (See Fig.~\ref{fig:app3} in Appendix C.)

Structures were predicted and analyzed regardless of whether their sequences appeared in either the AlphaFold3 training or validation datasets. To assess the potential influence of training-data leakage, we additionally constructed a filtered subset of the data containing only mutant sequences deposited after January 12, 2023, which is the cutoff date for the AlphaFold3 training and validation datasets. We further require that each mutant sequence has less than 100\% sequence identity to any sequence included in either the AlphaFold3 training or validation datasets prior to the cutoff date. This filtering procedure resulted in a subset of $181$ mutant sequences, for which the results were consistent with those obtained from the full dataset. (See Fig.~\ref{fig:app4} in Appendix D.) 

The average predicted mutant deformation $\bar{D}_{i}^{\mathrm{mut,pred}}$ is calculated analogously to the average mutant deformation $\bar{D}_{i}^{\mathrm{mut}}$ in Eq.~3, where $D_i$ is instead averaged over all pairs of experimental wild-type duplicate and predicted mutant structures for a given mutant sequence. The normalized predicted mutation-induced deformation is then
\begin{equation}
\widetilde{D}_{i}^{\mathrm{pred}} =
\frac{\bar{D}_{i}^{\mathrm{mut,pred}}}
{\bar{D}_{i}^{\mathrm{dup}}}.
\end{equation}

\subsection{Change in Relative Solvent Accessible Surface Area}

Changes in solvent accessibility have been shown to be correlated with changes in biophysical properties caused by mutations in proteins, particularly for mutations involving core residues~\cite{tsishyn2025,huiling2005}. To investigate the relationship between changes in solvent accessibility and local structural deformation upon mutation, we first calculated the solvent-accessible surface area (SASA) of each residue using the FreeSASA library~\cite{mitternacht2016} with the NACCESS implementation of a rolling-probe representation of the Lee--Richards molecular surface~\cite{lee1971,hubbard1993}. To normalize SASA across different residue types, we defined the relative solvent-accessible surface area (rSASA) as the ratio of the SASA of residue $i$ in the context of the protein structure $p$ to the SASA of the same residue in a Gly--X--Gly reference peptide, while preserving its backbone and side-chain dihedral angles:
\begin{equation}
\mathrm{rSASA}_{i}(p) =
\frac{\mathrm{SASA}_{i}^{\mathrm{context}}(p)}
{\mathrm{SASA}_{i}^{\mathrm{dipeptide}}(p)}.
\end{equation}

We quantify the average change in relative solvent accessibility of residue $i$ between wild-type duplicate structures $p$ and $q$ for a given wild-type sequence as the geometric mean of the absolute difference in $\mathrm{rSASA}_{i}(p)$ and $\mathrm{rSASA}_{i}(q)$:  
\begin{equation}
\hspace*{-0.8em}\scalebox{0.8}{$\displaystyle
\overline{\Delta \mathrm{rSASA}}_{i}^{\mathrm{dup}} = 
\left(\prod_{p=1}^{N_{d}-1} \prod_{q=p+1}^{N_{d}} \left| \mathrm{rSASA}_{i}(p)-\mathrm{rSASA}_{i}(q) \right| \right)^{\frac{2}{N_{d}(N_{d}-1)}}.
$}
\end{equation}
We use absolute differences between $\mathrm{rSASA}_{i}(p)$ and $\mathrm{rSASA}_{i}(q)$ rather than ratios because absolute differences are symmetric with respect to the two structures being compared. We use the geometric mean to calculate $\overline{\Delta \mathrm{rSASA}}_{i}^{\mathrm{dup}}$ because the nonzero absolute differences in rSASA span several orders of magnitude. When calculating the geometric mean, pairwise differences that equal to zero were excluded from the product. If all pairwise differences in $\mathrm{rSASA}_i$ were zero, corresponding to identical $\mathrm{rSASA}_i$ values across all structures, we set $\overline{\Delta \mathrm{rSASA}}_{i}^{\mathrm{dup}}=0$. Similarly, the average change in relative solvent accessibility of residue $i$ between wild-type duplicate--mutant structure pairs $p$ and $q$ for a given mutant sequence:
\begin{equation}
\hspace*{-0.8em}\scalebox{0.8}{$\displaystyle
\overline{\Delta \mathrm{rSASA}}_{i}^{\mathrm{mut}} = 
\left(\prod_{p=1}^{N_{m}} \prod_{q=1}^{N_{d}} \left| \mathrm{rSASA}_{i}(p)-\mathrm{rSASA}_{i}(q) \right| \right)^{\frac{1}{N_{m}N_{d}}}.
$}
\end{equation}
Finally, we define the normalized mutation-induced change in rSASA of residue $i$ as 
\begin{equation}
\widetilde{ \Delta \mathrm{rSASA}_{i}} = \frac{ \sum_{j=1}^{{\cal N}(i)}\overline{\Delta \mathrm{rSASA}_{j}}^{\text{ } \mathrm{ mut}}}{\sum_{j=1}^{{\cal N}(i)}\overline{\Delta \mathrm{rSASA}_{j}}^{\text{ } \mathrm{ \mathrm{dup}}}},
\end{equation}
where ${\cal N}(i)$ is the union of Voronoi neighbors of residue $i$ among all wild-type duplicate structures. $\overline{\Delta \mathrm{rSASA}_{j}}^{\mathrm{dup}}$ and $\overline{\Delta \mathrm{rSASA}_{j}}^{\mathrm{mut}}$ are summed over neighboring residues to capture local rearrangements. 

\begin{figure*}[t]
\centering
\includegraphics[width=0.85\linewidth]{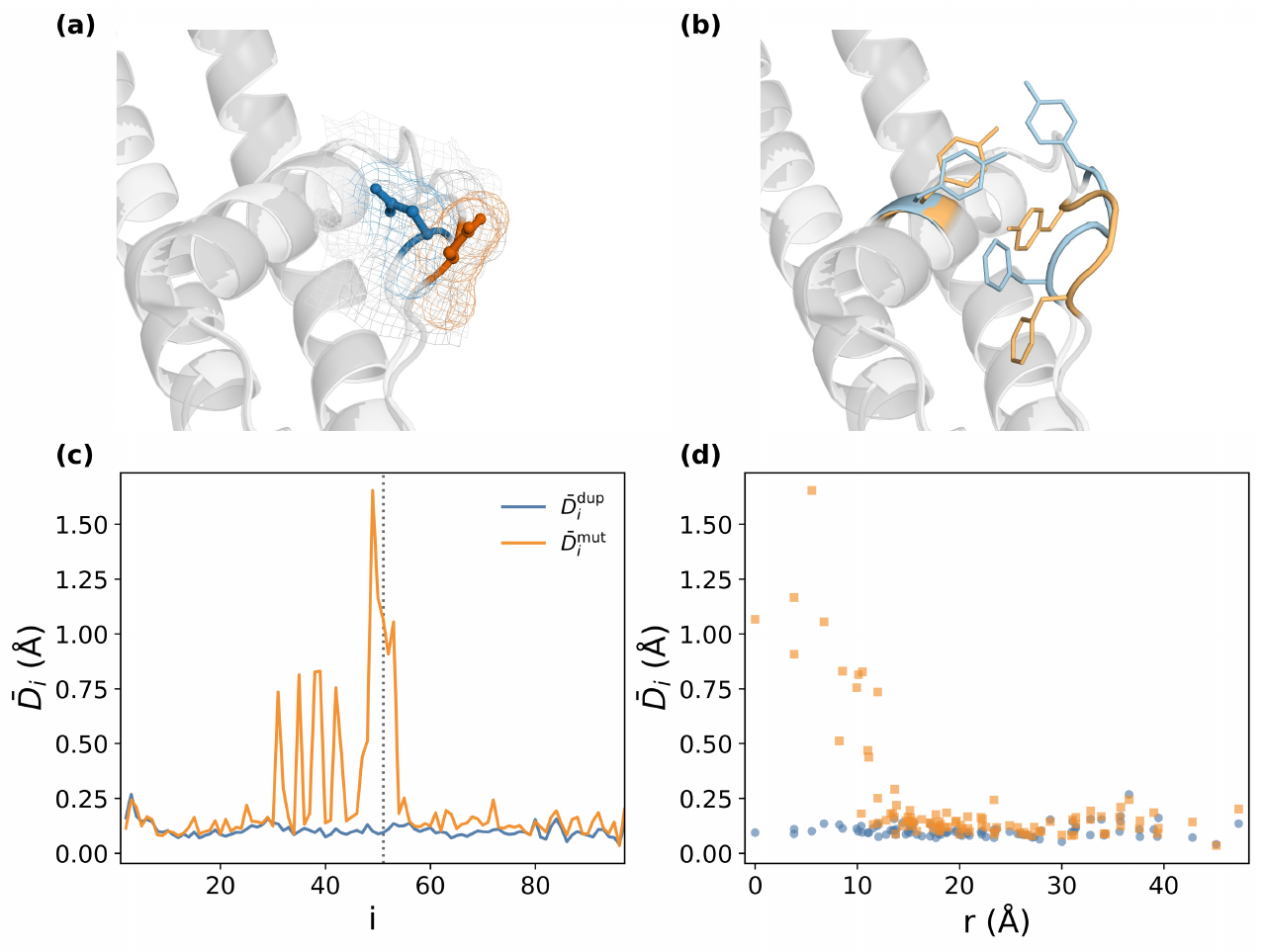}
\caption{
(a) Overlay of a representative wild-type duplicate structure of potassium selective channel protein (PDB ID: 4PDV) and a structure from a single point mutation, Asp-to-Asn at residue 51 (PDB ID: 3T1C). The mutation-site residues are shown in dark blue and dark orange for the wild-type duplicate and mutant structures, respectively. The mesh represents the volume occupied by residue 51 in the wild-type duplicate and mutant structures.
(b) Local structural rearrangement surrounding the mutation site. The backbone and side chains of neighboring residues are shown in light blue and light orange for the wild-type duplicate and mutant structures, respectively.
(c) Average duplicate deformation, $\bar{D}_i^{\mathrm{dup}}$ (blue) and average mutant deformation, $\bar{D}_i^{\mathrm{mut}}$ (orange) plotted as a function of residue $i$ for the wild-type--mutant sequence pair in (a). (d) Average duplicate deformation, $\bar{D}_i^{\mathrm{dup}}$ (blue) and average mutant deformation, $\bar{D}_i^{\mathrm{mut}}$ (orange) plotted as a function of the spatial distance $r$ between the C$_{\alpha}$ atom in each residue $i$ and the site of the mutation.
}
\label{fig:1}
\end{figure*}

\section{Results}

How should we quantify the structural changes in a protein caused by a single point mutation? As an illustrative example, we consider a potassium-selective channel protein (PDB ID: 4PDV) and one of its single point mutants~\cite{sauer2011} (PDB ID: 3T1C). Figure~\ref{fig:1}a shows an overlay of representative wild-type duplicate and mutant structures corresponding to the single point mutation Asp-to-Asn at residue 51 in the potassium selective channel protein. In Fig.~\ref{fig:1}b, we show the local environment surrounding the mutated residue in greater detail, highlighting changes in the protein backbone and conformations of side chains for neighboring residues. To quantify these structural changes, in Fig.~\ref{fig:1}c we calculate the local deformation for each residue $i$ between pairs of wild-type duplicate structures $\bar{D}_{i}^{\mathrm{dup}}$ and between wild-type duplicate and mutant structures  $\bar{D}_{i}^{\mathrm{mut}}$. The mutation-induced deformation is largest near the mutation site, whereas residues farther from the mutation site exhibit deformations comparable to the intrinsic structural fluctuations among wild-type duplicate structures. In Fig.~\ref{fig:1}d, we plot the same deformation measures as a function of the spatial distance between the C$_{\alpha}$ atom in each residue and the site of the mutation, showing that the mutation-induced deformation decreases with increasing distance from the mutation site.

\begin{figure*}[t]
\centering
\includegraphics[width=1\linewidth]{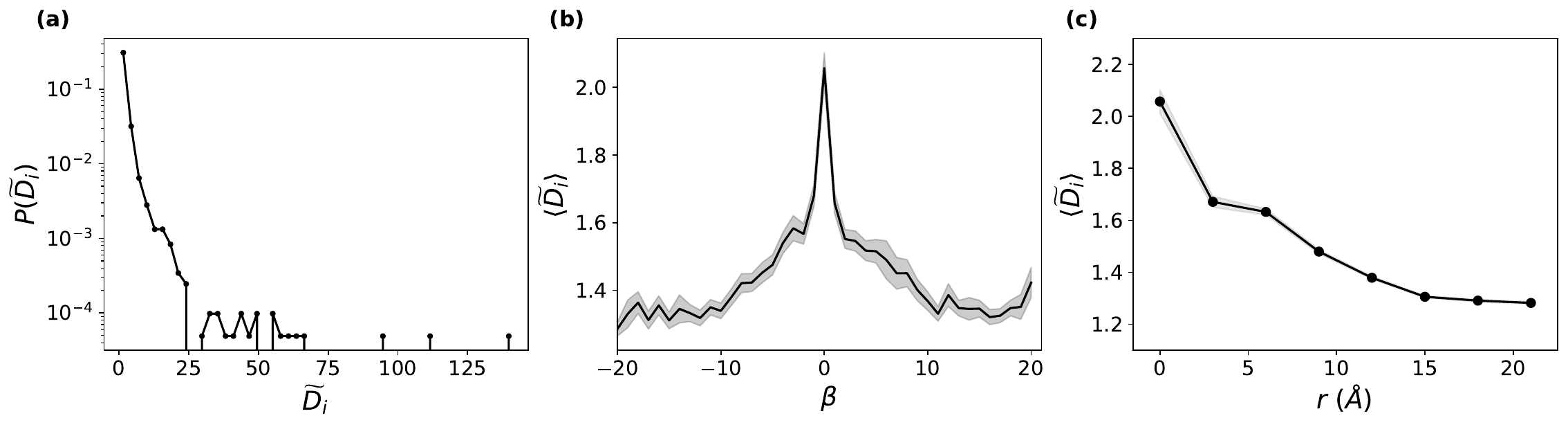}
\caption{
(a) Probability distribution of the normalized mutation-induced deformation, $\widetilde{D}_i$, at the mutation site across $N_s$ wild-type--mutant sequence pairs.
(b) Normalized mutation-induced deformation averaged over $N_s$ wildtype--mutant sequence pairs, $\langle \widetilde{D}_i \rangle$, plotted as a function of sequence distance $\beta \equiv i-m$ between the $i$th residue and the site of the mutation $m$. The shaded regions indicate the standard error of the mean across the sequence pairs.
(c) Normalized mutation-induced deformation averaged over $N_s$ wildtype--mutant sequence pairs, $\langle \widetilde{D}_i \rangle$, plotted as a function of distance $r \equiv |\vec{r}_m-\vec{r}_i|$. Residues are grouped into $3$~\AA\ distance bins according to their average distance from the mutation site in the wild-type duplicate structures. Shaded regions indicate the standard error of the mean within each distance bin.
}
\label{fig:2}
\end{figure*}

\subsection{Mutation-Induced Deformation Is Localized Near the Mutation Site}

In the previous section, we showed the observed deformation among wild-type duplicate structures $\bar{D}_{i}^{\mathrm{dup}}$ and the mutation-induced deformation $\bar{D}_{i}^{\mathrm{mut}}$ for a single wild-type--mutant sequence pair. We now expand this analysis by calculating the normalized mutation-induced deformation, $\widetilde{D}_{i}$, for all $6,967$ wild-type--mutant sequence pairs. As shown in Fig.~\ref{fig:2}a, $P({\widetilde D}_i)$ is roughly an exponential distribution. A substantial fraction of mutations have $\widetilde{D}_{i} \approx 1$ at the mutation site, indicating that for most single point mutations, the average mutant deformation is comparable to the intrinsic structural fluctuations among wild-type duplicate structures. This result is consistent with previous studies suggesting that many single-point mutations produce only limited structural perturbations~\cite{shakhnovich1991,guo2004}. A small fraction of mutations can produce substantially larger values of $\widetilde{D}_{i}$. Note that these large $\widetilde{D}_{i}$ are not restricted to residues with unusually small duplicate deformations but occur over a broad range of mutant and duplicate deformations (Fig.~\ref{fig:app2} in Appendix B). 

We show in Fig.~\ref{fig:2}b that the normalized mutation-induced deformation averaged over all wild-type--mutant sequence pairs, $\langle \widetilde{D}_{i}\rangle$, reaches a maximum value of approximately $2$ at the mutation site and decreases with increasing sequence distance $|\beta|$ from the mutation site. The deformation decreases by $\approx 30 \%$ within approximately $\pm 3$ residues and approaches a plateau value corresponding to the wildtype-duplicate fluctuations within $\pm 10$ residues. Similar behavior for $\langle \widetilde{D}_{i}(r)\rangle$ is observed as a function of spatial distance $r$ in Fig.~\ref{fig:2}c. $\langle \widetilde{D}_{i}(r)\rangle$ decreases substantially within approximately $3$~\AA\ from the mutation site, corresponding on average to the first shell of neighboring residues, and approaches a plateau value for $r \gtrsim 12$--$15$~\AA. Although $\langle \widetilde{D}_{i}\rangle$ remains slightly greater than 1 far from the mutation site in both sequence and spatial distance, indicating a small increase in deformation relative to the intrinsic structural fluctuations among wild-type duplicate structures, the strongest mutation-induced structural changes are concentrated near the mutation site.

\subsection{AlphaFold3 Prediction Accuracy Decreases for Strongly Perturbative Mutations}

\begin{figure*}[t]
\centering
\includegraphics[width=1\linewidth]{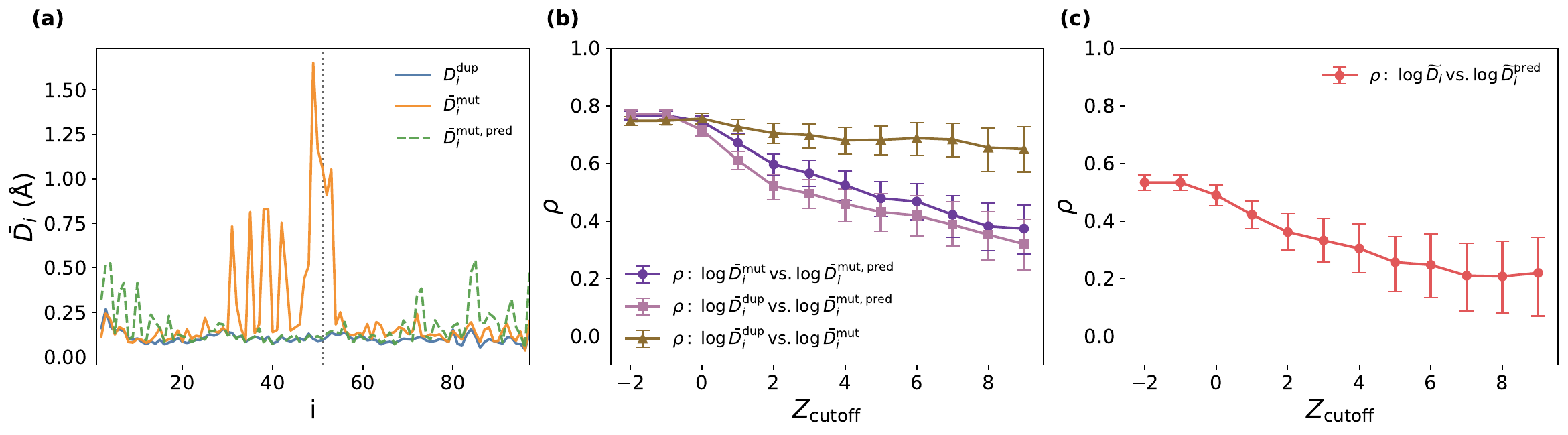}
\caption{
(a) Average duplicate deformation, $\bar{D}_i^{\mathrm{dup}}$ (blue), average mutant deformation, $\bar{D}_i^{\mathrm{mut}}$ (orange), and average predicted mutant deformation, $\bar{D}_i^{\mathrm{mut,pred}}$ (green), plotted as a function of residue index $i$ for a representative wild-type duplicate structure (PDB ID: 4PDV) and its single-point mutant structure (PDB ID: 3T1C), corresponding to an Asp-to-Asn substitution at residue 51.
(b) Pearson correlation coefficient $\rho$ between the average mutant deformation $\log\bar{D}_i^{\mathrm{mut}}$ and the average predicted mutant deformation $\log\bar{D}_i^{\mathrm{mut,pred}}$ (purple circles); between the average duplicate deformation $\log\bar{D}_i^{\mathrm{dup}}$ and the average predicted mutant deformation $\log\bar{D}_i^{\mathrm{mut,pred}}$ (pink squares); and between the average duplicate deformation $\log\bar{D}_i^{\mathrm{dup}}$ and the average mutant deformation $\log\bar{D}_i^{\mathrm{mut}}$ (brown triangles), computed over residues at mutation sites with $Z_i \geq Z_{\mathrm{cutoff}}$. Error bars indicate the 95\% confidence interval estimated from $10^3$ bootstrap resamples.
(c) Pearson correlation coefficient $\rho$ between the normalized mutation-induced deformation $\log \widetilde{D}_i$, and the normalized predicted mutation-induced deformation $\log\widetilde{D}_i^{\mathrm{pred}}$, computed over residues at mutation sites with $Z_i \geq Z_{\mathrm{cutoff}}$. Error bars indicate the 95\% confidence interval estimated from $10^3$ bootstrap resamples of the data.
}
\label{fig:3}
\end{figure*}

To evaluate how accurately AlphaFold3 captures mutation-induced structural deformation, we predicted all-atom structures for each mutant sequence in the dataset. We calculated the average predicted mutant deformation, $\bar{D}_i^{\mathrm{mut,pred}}$, by comparing the predicted mutant structures with the corresponding wild-type duplicate structures. Figure~\ref{fig:3}a shows a representative example in which the average mutant deformation, $\bar{D}_i^{\mathrm{mut}}$, is substantially larger than the average duplicate deformation, $\bar{D}_i^{\mathrm{dup}}$, near the mutation site, whereas the average predicted mutant deformation, $\bar{D}_i^{\mathrm{mut,pred}}$, remains close to $\bar{D}_i^{\mathrm{dup}}$ for all residues. This example suggests that AlphaFold3 predicts wild-type-like structures, even for mutant sequences, without capturing the deformation induced by the mutation. To test this behavior across the full dataset, we calculated the Pearson correlations between the average mutant deformation, $\log \bar{D}_i^{\mathrm{mut}}$, average predicted mutant deformation, $\log \bar{D}_i^{\mathrm{mut,pred}}$, and average duplicate deformation, $\log \bar{D}_i^{\mathrm{dup}}$, at the mutation site. To quantify the magnitude of the experimentally observed mutation-induced deformation, we used the $Z$-score, $Z_i$, defined in Eq. 5. In Fig.~\ref{fig:3}b, we plot the three pairwise Pearson correlations for mutation sites satisfying $Z_i \geq Z_{\mathrm{cutoff}}$ as a function of $Z_{\mathrm{cutoff}}$.

As shown in Fig.~\ref{fig:3}b, the Pearson correlation between the average mutant deformation, $\log \bar{D}_i^{\mathrm{mut}}$, and the average predicted mutant deformation, $\log\bar{D}_i^{\mathrm{mut,pred}}$, is $\rho \approx 0.75$ across all mutation-site residues (corresponding to $Z_{\mathrm{cutoff}}=-2$). However, this relatively high correlation does not necessarily indicate accurate prediction of mutation-induced structural changes. The correlation between $\log \bar{D}_i^{\mathrm{mut}}$ and the average duplicate deformation, $\log \bar{D}_i^{\mathrm{dup}}$, is comparably high, indicating that AlphaFold3 does not outperform the duplicate fluctuation baseline in predicting the mutant deformation. Moreover, $\log \bar{D}_i^{\mathrm{mut,pred}}$ is also strongly correlated with $\log \bar{D}_i^{\mathrm{dup}}$, suggesting that the predicted mutant structures resemble the duplicate structural ensemble near the mutation site rather than capturing mutation-induced deformation. As $Z_{\mathrm{cutoff}}$ increases, the correlation between $\log \bar{D}_i^{\mathrm{mut}}$ and $\log \bar{D}_i^{\mathrm{mut,pred}}$ decreases substantially, as does the correlation between $\log \bar{D}_i^{\mathrm{dup}}$ and $\log \bar{D}_i^{\mathrm{mut,pred}}$. In contrast, the Pearson correlation between $\log \bar{D}_i^{\mathrm{mut}}$ and $\log \bar{D}_i^{\mathrm{dup}}$ decreases only slightly with increasing $Z_{\mathrm{cutoff}}$. These results indicate that for single point mutations producing larger structural deformations, AlphaFold3 fails to recover the experimentally observed mutant deformation and the duplicate structural baseline. In addition, AlphaFold3 performs worse than using the duplicate fluctuations as the baseline for high-deformation mutations.

In Fig.~\ref{fig:3}c, we plot the Pearson correlation between the normalized mutation-induced deformation, $\log \widetilde{D}_i$, and the normalized predicted mutation-induced deformation, $\log \widetilde{D}_i^{\mathrm{pred}}$, as a function of $Z_{\mathrm{cutoff}}$. Normalizing by the average duplicate deformation removes the variation associated with intrinsic duplicate fluctuations to isolate the mutation-induced structural response. Across all mutation-site residues, the correlation is $\rho \approx 0.55$. However, for strongly perturbative mutations with $Z_{\mathrm{cutoff}} > 5$, the correlation decreases substantially to $\rho \approx 0.2$. Together, these results indicate that the putative high accuracy of AlphaFold3 predictions across the full wildtype-mutant dataset is caused by low $Z$-score mutations, for which the predicted deformation largely reflects the intrinsic fluctuations observed among duplicate structures. When the analysis is restricted to high $Z$-score, strongly perturbative mutations, AlphaFold3 predictions fail to accurately capture the mutation-induced structural deformation. This conclusion is robust to several possible confounding factors examined in Appendices D, E, and F. Restricting the analysis to high-confidence predictions with pLDDT $\geq 90$ does not improve the Pearson correlation at high $Z_{\mathrm{cutoff}}$ (Fig.~\ref{fig:app5} in Appendix E). More importantly, predictions for mutant sequences without an exact sequence match in the AlphaFold3 training data ($I_{\mathrm{seq}} < 100\%$) perform substantially worse and deteriorate more rapidly with increasing $Z_{\mathrm{cutoff}}$ (Fig.~\ref{fig:app4} in Appendix D). This result suggests that the performance reported for the full dataset may overestimate AlphaFold3's predictive accuracy for previously unseen mutant sequences. Moreover, the same deterioration in prediction accuracy is observed using alternative evaluation metrics, such as the Spearman correlation and both the mean absolute error and root-mean-squared error (Fig.~\ref{fig:app6} in Appendix F). Thus, the reduced prediction accuracy for strongly perturbative mutations cannot be attributed to low AlphaFold3 confidence, sequence-identity effects, or the particular choice of the accuracy metric, indicating a fundamental limitation in AlphaFold3's ability to capture the mutation-induced structural changes that exceed the intrinsic fluctuations among duplicate structures. In addition, we performed the same analysis using the protein structure prediction software platform, Boltz-2, which builds on the diffusion-based framework of AlphaFold3 with refinements to its model architecture and training procedure~\cite{wohlwend2024,passaro2025}.
Similar results are obtained using Boltz-2 (Fig.~\ref{fig:app7} in Appendix G), which shows that the limited accuracy in predicting mutation-induced structural deformation, particularly for strongly perturbative mutations, is not specific to AlphaFold3.

\begin{figure*}[t]
\centering
\includegraphics[width=1\linewidth]{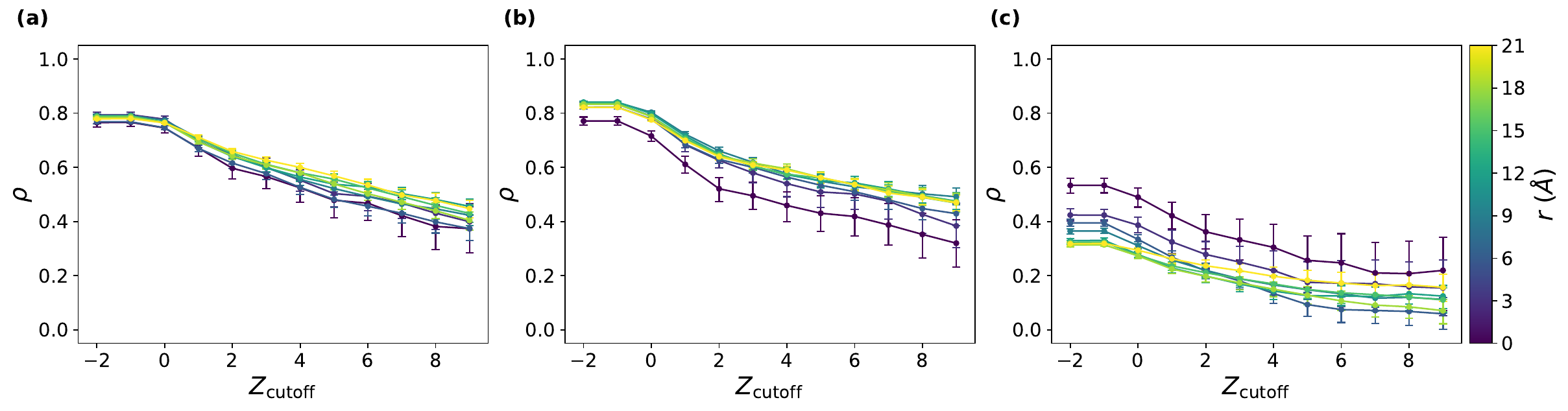}
\caption{
Pearson correlation coefficient between (a) the average mutant deformation, $\log \bar{D}_i^{\mathrm{mut}}$, and the average predicted mutant deformation, $\log \bar{D}_i^{\mathrm{mut,pred}}$; (b) the average duplicate deformation, $\log \bar{D}_i^{\mathrm{dup}}$, and the average predicted mutant deformation, $\log \bar{D}_i^{\mathrm{mut,pred}}$; and (c) the normalized mutation-induced deformation, $\log \widetilde{D}_i$, and the normalized predicted mutation-induced deformation, $\log \widetilde{D}_i^{\mathrm{pred}}$ plotted as a function of $Z_{\mathrm{cutoff}}$. For all panels, the color from blue to yellow indicates increasing distance $r$ from the mutation site. The error bars indicate the 95\% confidence intervals estimated from $10^3$ bootstrap resamples of the data.
}
\label{fig:4}
\end{figure*}

\subsection{Prediction Accuracy Depends Weakly on Distance From the Mutation Site}

Previous studies of mutant protein structure prediction~\cite{mcbride2023} found that AlphaFold2 predictions of mutation-induced structural changes correlated most strongly with experimentally observed structural changes at the mutation site, with the correlation decreasing with distance from the mutation site. We therefore extended our analysis from the mutation site to residues throughout the protein. In Fig.~\ref{fig:4}a--c, we plot the Pearson correlation between $\log \bar{D}_i^{\mathrm{mut}}(r)$ and $\log \bar{D}_i^{\mathrm{mut,pred}}(r)$, between $\log \bar{D}_i^{\mathrm{dup}}(r)$ and $\log \bar{D}_i^{\mathrm{mut,pred}}(r)$, and between $\log \widetilde{D}_i(r)$ and $\log \widetilde{D}_i^{\mathrm{pred}}(r)$, respectively, as a function of $Z_{\mathrm{cutoff}}$ and distance $r$ from the site of the mutation.

As shown in Fig.~\ref{fig:4}a and Fig.~\ref{fig:4}b, both the Pearson correlation between $\log \bar{D}_i^{\mathrm{mut}}(r)$ and $\log \bar{D}_i^{\mathrm{mut,pred}}(r)$ and the correlation between $\log \bar{D}_i^{\mathrm{dup}}(r)$ and $\log \bar{D}_i^{\mathrm{mut,pred}}(r)$ increase slightly with distance from the mutation site. In contrast, Fig.~\ref{fig:4}c shows that the correlation between $\log \widetilde{D}_i(r)$ and $\log \widetilde{D}_i^{\mathrm{pred}}(r)$ is greatest at the mutation site and decreases slightly with increasing $r$, consistent with previous results~\cite{mcbride2023}. Across all mutations, this correlation decreases from $\rho \approx 0.55$ at $r=0$ to $\rho \approx 0.35$ at large $r$. Thus, farther from the mutation site, the ability of AlphaFold3 to capture the normalized mutation-induced deformation decreases, while the predicted deformation becomes slightly more correlated with the deformation among wild-type duplicate structures. However, the dependence of the prediction accuracy on the distance from the site of the mutation is small compared to the pronounced decrease in correlation with increasing $Z_{\mathrm{cutoff}}$.

\subsection{Changes in Solvent Accessibility Identify Strongly Perturbative Mutations}

Since it is difficult for AlphaFold3 to predict strongly perturbative mutations, it is important to identify physical features that are associated with mutation-induced deformation. Because changes in solvent accessibility have previously been shown to be correlated with changes in biophysical properties caused by mutations in proteins~\cite{tsishyn2025,huiling2005}, we calculated the relative solvent-accessible surface area (rSASA) of each residue and quantified the change in rSASA between the mutant and wild-type duplicate structures relative to the intrinsic variations in rSASA among duplicate structures. We then summed these changes over the local neighborhood of the mutation site to define the normalized mutation-induced change in solvent accessibility, $\widetilde{\Delta \mathrm{rSASA}}_i$, as described in Eq.~11.

\begin{figure*}[!htbp]
\centering
\includegraphics[width=1\linewidth]{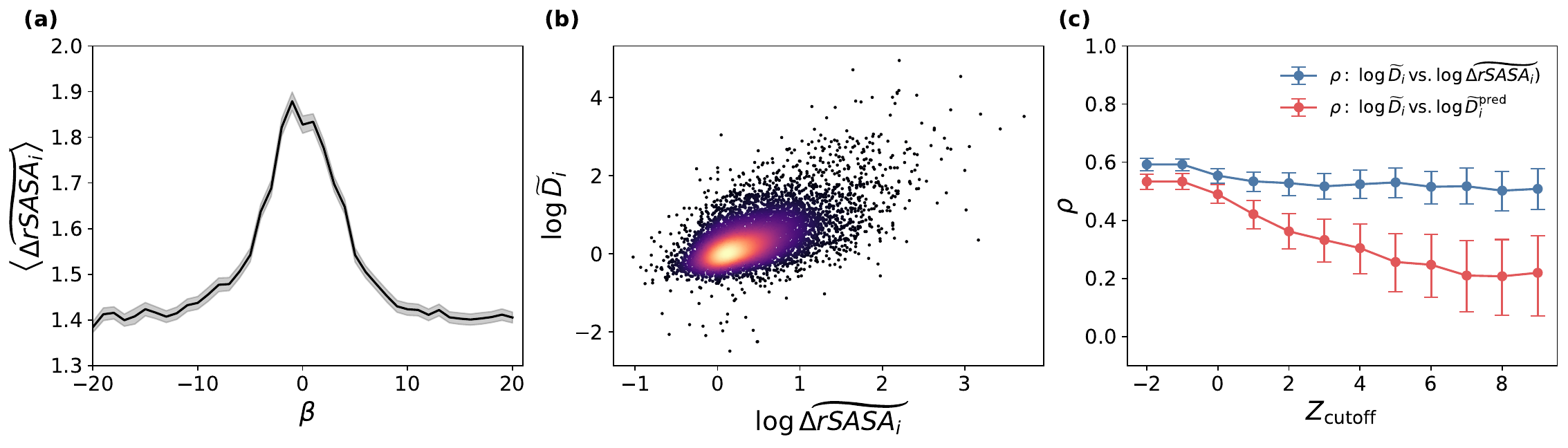}
\caption{
(a) Normalized change in relative solvent accessible surface area averaged over the $N_s$ wildtype-mutant sequence pairs, $\langle \widetilde{\Delta \mathrm{rSASA}}_i \rangle$, plotted as a function of sequence distance from the mutation site, $\beta$. Shaded regions indicate the standard error of the mean across the $N_s$ sequence pairs.
(b) Probability distribution (increasing from dark black to light yellow) of the normalized mutation-induced deformation, $\log \widetilde{D}_i$, plotted versus the normalized change in relative solvent accessible surface area, $\widetilde{\Delta \mathrm{rSASA}}_i$, at the mutation sites. 
(c) Pearson correlation coefficient between the normalized mutation-induced deformation, $\log \widetilde{D}_i$, and the normalized predicted mutation-induced deformation $\log\widetilde{D}_i^{\mathrm{pred}}$ (red); and between the normalized mutation-induced deformation $\log\widetilde{D}_i$, and the normalized change in relative solvent accessible surface area $\widetilde{\Delta \mathrm{rSASA}}_i$ (blue), computed over residues at mutation sites with $Z_i \geq Z_{\mathrm{cutoff}}$. The error bars indicate the 95\% confidence interval estimated from $10^3$ bootstrap resamples.
}
\label{fig:5}
\end{figure*}

In Fig.~\ref{fig:5}a, we plot $\widetilde{\Delta \mathrm{rSASA}}_i$ as a function of sequence distance from the mutation site, $\beta$. Similar to the normalized local deformation, $\widetilde{D}_i$, $\widetilde{\Delta \mathrm{rSASA}}_i$ exhibits a pronounced peak at the mutation site ($\beta=0$) and decreases with increasing sequence distance. This similarity between $\widetilde{D}_i$ and $\widetilde{\Delta \mathrm{rSASA}}_i$ emphasizes that mutation-induced changes in solvent accessibility are spatially localized near the mutation site, consistent with the localization of structural deformation. In Fig.~\ref{fig:5}b, we compare $\widetilde{\Delta \mathrm{rSASA}}_i$ and $\widetilde{D}_i$ at the mutation site across the full dataset. We find a Pearson correlation coefficient of $\rho \approx 0.6$ between $\log \widetilde{\Delta \mathrm{rSASA}}_i(0)$ and $\log \widetilde{D}_i(0)$, indicating a strong association between mutation-induced changes in solvent accessibility and local structural deformation.

To determine whether this relationship persists for strongly perturbative mutations, in Fig.~\ref{fig:5}c we compare the correlation between $\log \widetilde{\Delta \mathrm{rSASA}}_i(0)$ and $\log \widetilde{D}_i(0)$ with that between $\log \widetilde{D}_i^{\mathrm{pred}}(0)$ and $\log \widetilde{D}_i(0)$ as a function of $Z_{\mathrm{cutoff}}$. At $Z_{\mathrm{cutoff}}=-2$, which includes the full dataset, the two Pearson correlation coefficients are comparable, with the rSASA-based correlation slightly higher than that obtained from AlphaFold3-predicted structures. As $Z_{\mathrm{cutoff}}$ increases, the correlation between $\widetilde{\Delta \mathrm{rSASA}}_i$ and $\widetilde{D}_i$ remains nearly constant, whereas the correlation between AlphaFold3-predicted deformation and the experimentally observed deformation decreases significantly. Thus, changes in solvent accessibility remain strongly associated with structural deformation even for mutations that produce large structural perturbations. Because $\widetilde{\Delta \mathrm{rSASA}}_i$ is calculated using experimentally observed mutant structures, it cannot be used directly as a predictive feature when the mutant structure is unknown. Instead, $\widetilde{\Delta \mathrm{rSASA}}_i$ provides an approximate upper bound on the information that is needed to accurately predict mutation-induced deformation. These results suggest that improving predictions of mutation-induced changes in solvent accessibility can in turn, enhance the prediction accuracy for strongly perturbative mutations.

\section{Discussion}
\label{discussion}

In this article, we presented a framework for quantifying local mutation-induced structural deformation using a curated dataset of wildtype--mutant sequence pairs and their X-ray crystal structures. We quantify the local deformation, $D_i$, of each residue as the root-mean-square change in the C$_\alpha$ distances between residue $i$ and its neighboring residues to eliminate the effects of large-scale rigid-body motion. We then define the normalized mutation-induced deformation, $\widetilde{D}_i$, as the ratio of the average mutant deformation, $\bar{D}_i^{\mathrm{mut}}$, to the average deformation among wild-type duplicate structures, $\bar{D}_i^{\mathrm{dup}}$. Note that experimentally determined X-ray crystal structures more frequently contain mutations in flexible, solvent-exposed regions with higher $\bar{D}_i^{\mathrm{dup}}$ (Fig.~\ref{fig:app1} in Appendix A). Therefore, the normalization allows us to distinguish mutation-induced structural changes from the intrinsic structural fluctuations observed among wild-type duplicate structures. Applying this analysis to our dataset of nearly $7,000$ wildtype--mutant sequence pairs, we find that the distribution $P(\widetilde{D}_i)$ at the mutation site is approximately exponential, with a substantial fraction of mutations yielding $\widetilde{D}_i \approx 1$ and a much smaller fraction producing much larger values of $\widetilde{D}_i$. The observed exponential distribution suggests that single point mutations are independent events and agrees with previous results indicating that amino acid mutations on average do not result in significant structural changes~\cite{shakhnovich1991,guo2004}. We further find that $\widetilde{D}_i$ is strongly localized near the mutation site, decreasing by $30\%$ at $|\beta| \approx 3$ and $r \approx 3$~\AA, and further decaying to a plateau value of $\gtrsim 1$ for $|\beta| > 10$ and $r > 12$--$15\text{\AA}$.

We also evaluated the ability of AlphaFold3 to reproduce mutation-induced structural deformation. We first calculated the average predicted mutant deformation, $\bar{D}_i^{\mathrm{mut,pred}}$, analogously to $\bar{D}_i^{\mathrm{mut}}$ by comparing AlphaFold3-predicted mutant structures with the experimentally observed wild-type duplicate structures. Across all mutation sites, $\log \bar{D}_i^{\mathrm{mut,pred}}$ and $\log \bar{D}_i^{\mathrm{mut}}$ are strongly correlated, with Pearson correlation $\rho \approx 0.75$. However, the relatively high correlation in the unnormalized deformation does not necessarily indicate that AlphaFold3 captures the structural response to mutations. Since $\bar{D}_i^{\mathrm{mut}}$ is also strongly correlated with $\bar{D}_i^{\mathrm{dup}}$ it is necessary to define $\widetilde{D}_i^{\mathrm{pred}}$ by normalizing $\bar{D}_i^{\mathrm{mut,pred}}$ by $\bar{D}_i^{\mathrm{dup}}$ to evaluate how well mutation-induced deformation is predicted relative to the native structural fluctuations. We find that the correlation between $\log \widetilde{D}_i^{\mathrm{pred}}$ and $\log \widetilde{D}_i$ is approximately $\rho \approx 0.55$ and as we restrict the analysis to mutations with larger deformations $Z_{i} \geq Z_{\mathrm{cutoff}}$, the correlation between $\log \widetilde{D}_i^{\mathrm{pred}}$ and $\log \widetilde{D}_i$ decreases to $\rho \approx 0.2$. In contrast, the dependence on spatial distance from the mutation site is comparatively weak: across all mutations, the correlation decreases from $\rho \approx 0.55$ at $r=0$ to $\rho \approx 0.35$ at large $r$. Thus, the magnitude of the mutation-induced perturbation, rather than the distance from the mutation site, is the stronger determinant of AlphaFold3 prediction accuracy. In addition, mutant sequences without exact sequence matches in the training set show substantially weaker correlations with mutation-induced deformation, suggesting that the AlphaFold3 accuracy on the full dataset may be an overestimate for previously unseen mutant sequences. These results point to an important distinction between predicting the overall folded structure of a protein and predicting its structural response to a specific mutation. AlphaFold3 is trained primarily to reproduce a single experimentally observed protein structure for a given amino acid sequence~\cite{chakravarty2022} rather than to learn structural differences between matched wild-type and mutant proteins. Because single-point mutant structures, particularly paired wild-type duplicate-point mutant structures, are comparatively limited in the PDB, predictions for single-point mutants remain biased toward structural states represented by wild-type proteins. This bias reduces the sensitivity to single point mutations that induce structural changes that are substantially larger than the intrinsic wildtype-duplicate fluctuations.

The relationship between relative solvent accessibility and local deformation further highlights the value of physically interpretable descriptors for characterizing mutation-induced structural responses. We defined the normalized change in relative solvent accessibility, $\widetilde{\Delta \mathrm{rSASA}}_i$, from the mutation-induced change in rSASA within the local neighborhood of residue $i$, normalized by the corresponding variation among wild-type duplicate structures. Similar to $\widetilde{D}_i$, $\widetilde{\Delta \mathrm{rSASA}}_i$ is strongly localized near the mutation site with strong Pearson correlation ($\rho \approx 0.6$) between $\log \widetilde{\Delta \mathrm{rSASA}}_i$ and $\log \widetilde{D}_i$. More importantly, this correlation remains nearly constant as $Z_{\mathrm{cutoff}}$ increases, whereas the correlation between $\log \widetilde{D}_i^{\mathrm{pred}}$ and $\log \widetilde{D}_i$ decreases substantially. Thus, changes in solvent accessibility remain strongly associated with structural deformation even for strongly perturbative mutations. Because $\widetilde{\Delta \mathrm{rSASA}}_i$ is calculated using experimentally observed mutant structures, it cannot itself serve as a predictive feature when the mutant structure is unknown. Instead, the observed correlation can be interpreted as an approximate upper bound on the information that is required for  accurately predicting large mutation-induced changes in structure. Note that structural deformation does not require changes in solvent accessibility, e.g. when the mutant structure does not undergo changes in volume. Thus, the strong correlation between $\widetilde{\Delta \mathrm{rSASA}}_i$ and $\widetilde{D}_i$ is nontrivial and indicates that changes in local solvent exposure capture a key feature of mutation-induced structural rearrangements. The persistence of this relationship for strongly deforming mutations suggests that relatively simple physical properties can retain mutation-specific structural information that is not consistently captured by current large language model structure-prediction methods. We also find that the results remain qualitatively unchanged when mutations are divided into four classes (core-to-core, core-to-surface, surface-to-core, and surface-to-surface) based on whether the mutation site is located in the protein core or on the surface in the wild-type and mutant structures.

Several promising research directions can improve the prediction and mechanistic understanding of mutation-induced structural responses. Prediction methods can be developed specifically for predicting how a mutation changes a structure, starting from the wildtype structure, rather than independently predicting how the wild-type and mutant structures fold from unfolded states. Training on paired wildtype--mutant structures and incorporating objectives that emphasize mutation-induced differences may increase the sensitivity to structural changes that are not captured by conventional folding studies. A complementary direction would be to develop models that predict mutation-induced changes in physical features such as rSASA directly from sequence\cite{grigas2026}, especially for improving predictions of strongly perturbative mutations. In parallel, molecular dynamics (MD) simulations can provide a complementary framework for quantifying mutation-induced structural response. While most previous studies have focused on simulations of only the mutant protein structures~\cite{fukuyoshi2016,morra2008,abhinand2016}, an alternative approach would be to use MD simulations to directly quantify the structural response to a mutation relative to the intrinsic fluctuations of the wild-type protein. An equilibrium wild-type ensemble could first be simulated to characterize the intrinsic fluctuations. Mutations can then be introduced into the sampled wildtype structures, and the resulting structural response can be monitored as a function of time after the mutation was made as the system relaxes toward the `equilibrium' mutant structure. Comparing the wild-type and sampled mutant response structural ensembles will allow researchers to separate mutation-induced deformation from intrinsic thermal fluctuations of the wildtype structures.

\begin{acknowledgments}
The authors acknowledge support from NIH Training Grant T32GM145452, as well as the High Performance Computing facilities operated by Yale’s Center for Research Computing.
\end{acknowledgments}

\section*{Conflict of interest statement}
The authors declare no conflicts of interests.

\section*{Data availability statement}
Datasets and all code used for analysis can be found at \url{https://github.com/lzyttxs/}

\appendix
\renewcommand{\appendixname}{APPENDIX}

\section{Mutations Are More Likely to Occur in Solvent-Exposed Regions}
\label{app:a}

\begin{figure}[!htbp]
\centering
\includegraphics[width=1\linewidth]{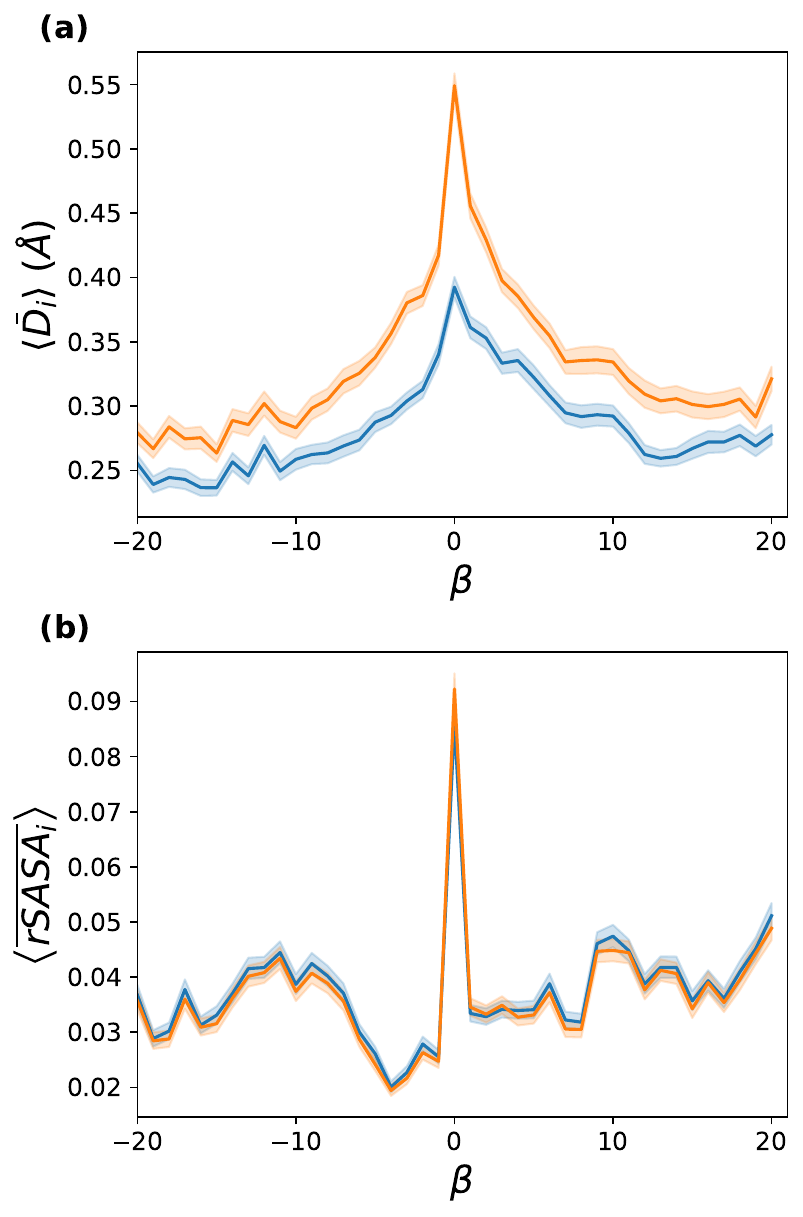}
\caption{
(a) Average duplicate deformation (blue) and average mutant deformation (orange), averaged over $N_s$ wildtype--mutant sequence pairs, $\langle \bar{D}_i \rangle$, plotted as a function of sequence distance from the mutation site, $\beta$. Error bars indicate the standard error of the mean across the $N_s$ sequence pairs.
(b) Average relative solvent-accessible surface area $\langle \overline{\mathrm{rSASA}}_i \rangle$ for wiltype-duplicate structures (blue) and mutant structures (orange), averaged over $N_s$ wildtype--mutant sequence pairs plotted as a function of sequence distance from the mutation site, $\beta$. Error bars indicate the standard error of the mean across the $N_s$ sequence pairs.
}
\label{fig:app1}
\end{figure}

In this appendix, we show that experimentally characterized mutation sites tend to occur in more solvent-exposed and fluctuation-prone regions of protein structures. We first show that the local deformation among wild-type duplicate structures is enhanced near the mutation site. As shown in Fig.~\ref{fig:app1}a, both $\langle \bar{D}_i^{\mathrm{dup}} \rangle$ and $\langle \bar{D}_i^{\mathrm{mut}} \rangle$, averaged over $N_s$ wildtype-mutant sequence pairs, possess a peak at the mutation site ($\beta=0$) and decrease with sequence distance from the mutation site. To examine whether this trend is associated with solvent accessibility, we calculate the average rSASA for a given wildtype sequence as
\begin{equation}
\overline{\mathrm{rSASA}}_i^{\mathrm{dup}} = \left(\prod_{p=1}^{N_d}\mathrm{rSASA}_i(p)\right)^{1/N_d},
\end{equation}
where $p$ is a protein structure in the wildtype-duplicate ensemble and $N_d$ is the total number of duplicate structures for a given wild-type sequence. Similarly, for a given mutant sequence,
\begin{equation}
\overline{\mathrm{rSASA}}_i^{\mathrm{mut}} = \left(\prod_{q=1}^{N_m}\mathrm{rSASA}_i(q)\right)^{1/N_m},
\end{equation}
where $q$ is a protein structure in the mutant structural ensemble and $N_m$ is the total number of mutant structures for a given wildtype-mutant sequence pair. As shown in Fig.~\ref{fig:app1}b, both $\langle \overline{\mathrm{rSASA}}_i^{\mathrm{dup}} \rangle$ and $\langle \overline{\mathrm{rSASA}}_i^{\mathrm{mut}} \rangle$, averaged over $N_s$ wildtype--mutant sequence pairs, also peak at the mutation site $\beta=0$. These results indicate that single point mutations for which both wild-type and mutant x-ray crystal structures are available are preferentially located in more solvent-exposed regions. The elevated $\langle \bar{D}_i^{\mathrm{dup}} \rangle$ at these sites is therefore consistent with greater intrinsic structural variability in the regions where experimentally characterized mutations occur. These results further motivate the use of the wild-type duplicate deformation, $\bar{D}_i^{\mathrm{dup}}$, as a baseline for normalizing mutation-induced deformation.

\section{Relationship Between Duplicate, Mutant and Normalized Deformations}

\label{app:b}

\begin{figure*}[!htbp]
\centering
\includegraphics[width=1\linewidth]{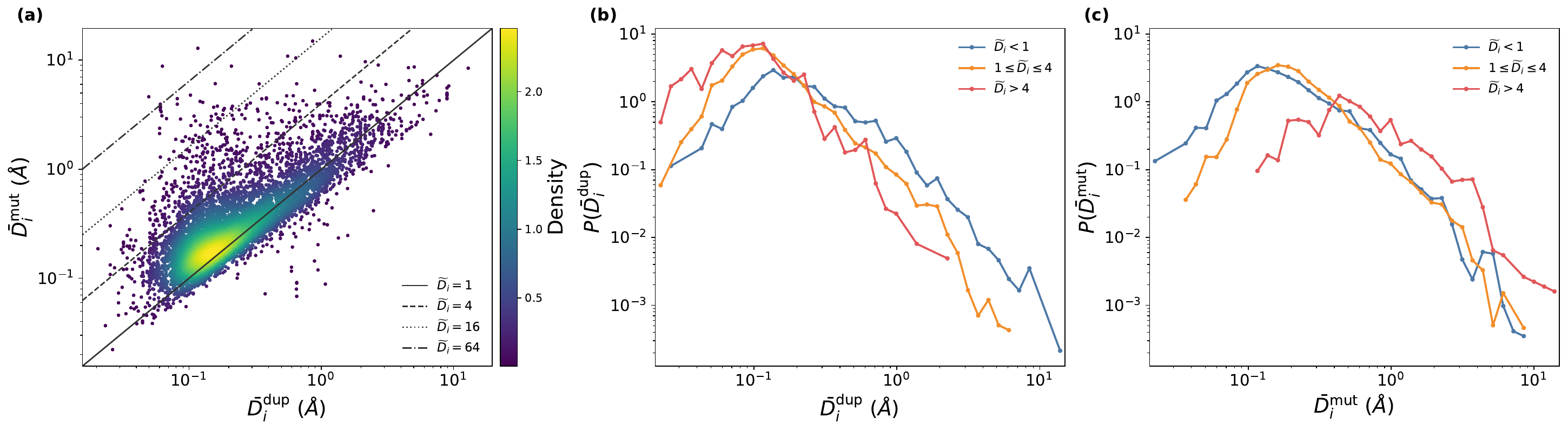}
\caption{
(a) Average mutant deformation $\bar D_i^{\mathrm{mut}}$ plotted versus average duplicate deformation $\bar D_i^{\mathrm{dup}}$, at the mutation site for $N_s$ experimental wild-type--mutant sequence pairs. Each point represents one pair and is colored according to the local density. Diagonal lines indicate constant normalized mutation-induced deformation, $\widetilde D_i=1,4,16,$ and $64$.
(b,c) Probability distributions of (b) $\bar D_i^{\mathrm{dup}}$ and (c) $\bar D_i^{\mathrm{mut}}$ for three groups of sequence pairs classified by their normalized deformation: $\widetilde D_i<1$, $1\leq\widetilde D_i\leq4$, and $\widetilde D_i>4$ in logarithmically spaced bins.}
\label{fig:app2}
\end{figure*}

In this appendix, we investigate whether large normalized mutation-induced deformations arise primarily from small intrinsic structural variability. We examine the relationship between mean mutant deformation $\bar D_i^{\mathrm{mut}}$, mean duplicate deformation $\bar D_i^{\mathrm{dup}}$ and normalized mutation-induced deformation $\widetilde D_i$ at the mutation site. In Fig.~\ref{fig:app2}a, we find that $\bar D_i^{\mathrm{mut}}$ and $\bar D_i^{\mathrm{dup}}$ are positively correlated, with most sequence pairs occuring near or slightly above the diagonal $\widetilde D_i=1$. Sequence pairs with large normalized deformations lie above this diagonal but span a broad range of both $\bar D_i^{\mathrm{mut}}$ and $\bar D_i^{\mathrm{dup}}$. To further investigate this relationship, we compared the distributions of $\bar D_i^{\mathrm{dup}}$ and $\bar D_i^{\mathrm{mut}}$ for sequence pairs with $\widetilde D_i<1$, $1\leq\widetilde D_i\leq4$, and $\widetilde D_i>4$ in Fig.~\ref{fig:app2}b and c. The distributions of $\bar D_i^{\mathrm{dup}}$ show a slight shift toward smaller values as $\widetilde D_i$ increases, while the distributions of $\bar D_i^{\mathrm{mut}}$ exhibit a pronounced shift toward larger values with increasing $\widetilde D_i$. These results emphasize that large normalized mutation-induced deformations reflect the magnitudes of average mutant and duplicate deformations.

\section{Robustness of Predicted Structural Deformation to the Number of AlphaFold3 Predictions}
\label{app:c}

\begin{figure*}[!htbp]
\centering
\includegraphics[width=1\linewidth]{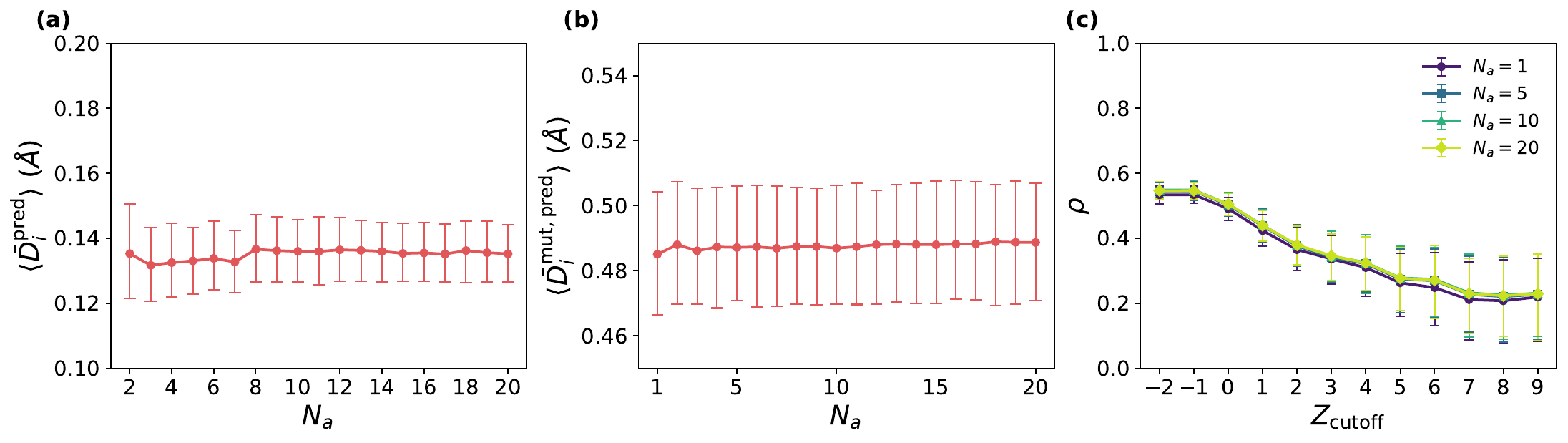}
\caption{
(a) Average predicted variability, $\langle \bar{D}_i^{\mathrm{pred}} \rangle$, at the mutation site $i$, averaged over $N_s$ wild-type--mutant sequence pairs, plotted as a function of the number of predicted mutant structures, $N_a$, generated from different random seeds. Error bars indicate the 95\% confidence intervals estimated from $10^3$ bootstrap resamples.
(b) Average predicted mutant deformation, $\langle \bar{D}_i^{\mathrm{mut,pred}} \rangle$, at the mutation site $i$, averaged over $N_s$ wild-type--mutant sequence pairs, plotted as a function of the number of predicted mutant structures, $N_a$. Error bars indicate the 95\% confidence intervals estimated from $10^3$ bootstrap resamples.
(c) Pearson correlation coefficient between $\log \widetilde{D}_i$ and $\log \widetilde{D}_i^{\mathrm{pred}}$ at the mutation site $i$, calculated using $N_a$ predicted mutant structures, plotted as a function of $Z_{\mathrm{cutoff}}$ for mutation sites satisfying $Z_i \geq Z_{\mathrm{cutoff}}$. Error bars indicate the 95\% confidence intervals estimated from $10^3$ bootstrap resamples.
}
\label{fig:app3}
\end{figure*}

We further tested whether the prediction accuracy depends on the number of AlphaFold3-predicted mutant structures generated using different random seeds. To quantify the structural variability among the predicted mutant structures, we define the variability in the predicted deformation at residue $i$ for a given mutant sequence as
\begin{equation}
\bar{D}_i^{\mathrm{pred}} =
\frac{2}{N_a(N_a-1)}
\sum_{p=1}^{N_a-1}\sum_{q=p+1}^{N_a} D_i(p,q),
\end{equation}
where $N_a$ is the number of predicted mutant structures for a given mutant sequence and the summation runs over all unique pairs of predicted mutant structures. We varied $N_a$ by including predicted structures generated from different numbers of random seeds. The average predicted variability at the mutation site $i$, $\bar{D}_i^{\mathrm{pred}}$, remains nearly constant as $N_a$ increases (Fig.~\ref{fig:app3}a), indicating that the structural variability among AlphaFold3 predictions is insensitive to the number of random seeds used. Similarly, the average predicted mutant deformation $\bar{D}_i^{\mathrm{mut,pred}}$, shows little dependence on $N_a$ (Fig.~\ref{fig:app3}b). Finally, the Pearson correlation between the normalized mutation-induced deformation, $\log \widetilde{D}_i$, and the normalized predicted mutation-induced deformation, $\log \widetilde{D}_i^{\mathrm{pred}}$, remains nearly constant as a function of $N_a$ over the full range of $Z_{\mathrm{cutoff}}$ (Fig.~\ref{fig:app3}c). Thus, increasing the number of predicted structures generated from different random seeds neither changes the predicted deformation nor improves its agreement with experimentally observed structures, indicating that our conclusions are robust to the number of AlphaFold3 random seeds used.

\section{Effect of Sequence Identity on AlphaFold3 Prediction Accuracy}
\label{app:d}

\begin{figure}[!htbp]
\centering
\includegraphics[width=1\linewidth]{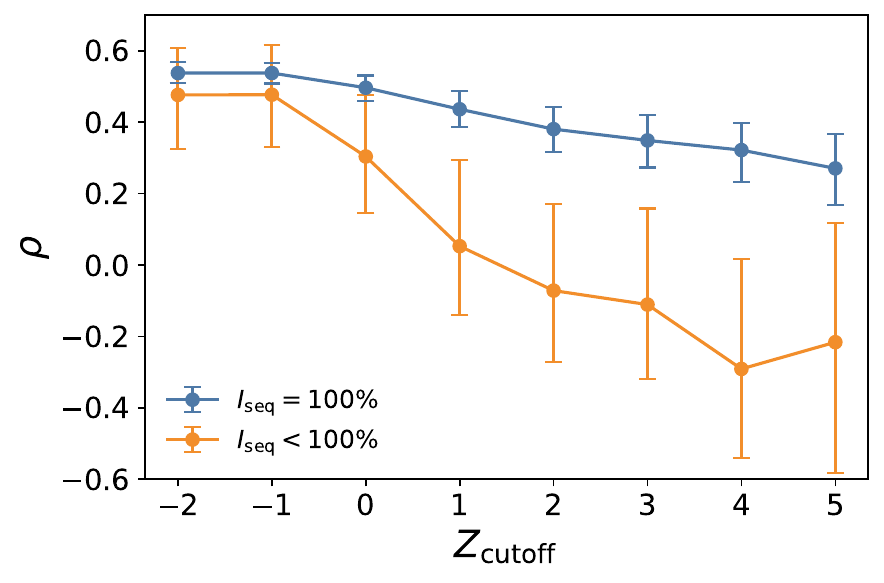}
\caption{
Pearson correlation coefficient between the normalized mutation-induced deformation $\log \widetilde{D}_i$, and the normalized predicted mutation-induced deformation, $\log \widetilde{D}_i^{\mathrm{pred}}$, at mutation sites for mutant sequences with sequence identity $I_{\mathrm{seq}}=100\%$ (blue) and $I_{\mathrm{seq}}<100\%$ (orange), calculated for residues satisfying $Z_i \geq Z_{\mathrm{cutoff}}$. The error bars indicate the 95\% confidence interval estimated from $10^3$ bootstrap resamples.
}
\label{fig:app4}
\end{figure}

We next examine whether AlphaFold3's ability to predict mutation-induced structural deformation depends on the sequence identity of the mutant sequence, $I_{\mathrm{seq}}$, to sequences represented in the AlphaFold3 training dataset. Figure~\ref{fig:app4} compares the Pearson correlation between the normalized mutation-induced deformation, $\log \widetilde{D}_i$, and the normalized predicted mutation-induced deformation, $\log \widetilde{D}_i^{\mathrm{pred}}$, for mutant sequences with $I_{\mathrm{seq}} = 100\%$ and $I_{\mathrm{seq}} < 100\%$. For mutant sequences without an exact sequence match ($I_{\mathrm{seq}} < 100\%$), the prediction accuracy is substantially worse than for mutant sequences with an exact sequence match: the correlation is consistently lower, decreases more rapidly with increasing $Z_{\mathrm{cutoff}}$, and even becomes negative for large $Z_{\mathrm{cutoff}}$. Although the $I_{\mathrm{seq}} < 100\%$ subset contains fewer data points (only 181 mutant sequences) and therefore has larger bootstrap uncertainties, the error bars for the two groups remain non-overlapping. These results indicate that the conclusions of the main-text analysis are not driven by poor performance within a particular sequence-identity subset. Rather, the prediction of mutation-induced deformation is already limited for mutant sequences with exact sequence matches and becomes even less accurate for sequences without such matches. Thus, the performance reported for the full dataset of $N_s$ wildtype--mutant sequence pairs may provide an inflated estimate of AlphaFold3's predictive accuracy for mutant sequences without exact matches in its training data.

\section{Effect of pLDDT on AlphaFold3 Prediction Accuracy}
\label{app:e}

\begin{figure}[!htbp]
\centering
\includegraphics[width=1\linewidth]{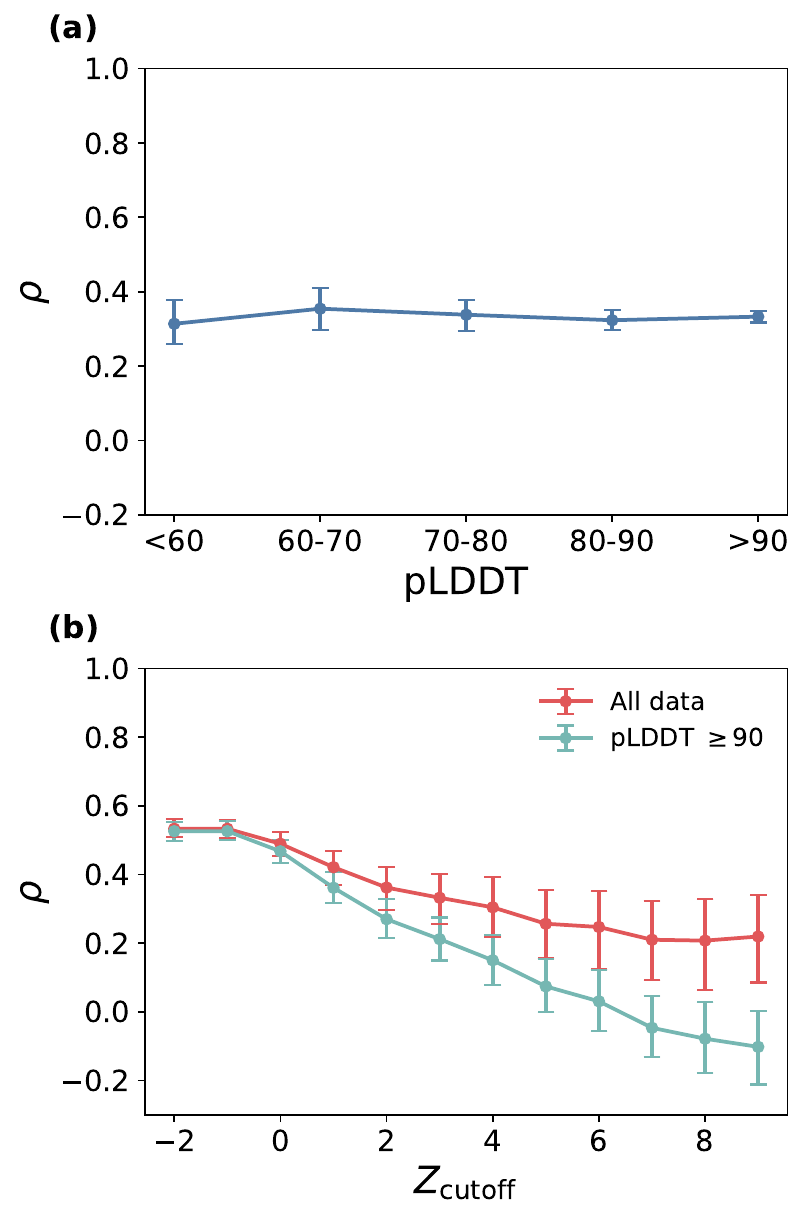}
\caption{
(a) Pearson correlation coefficient between the normalized mutation-induced deformation, $\log \widetilde{D}_i$, and the normalized predicted mutation-induced deformation, $\log \widetilde{D}_i^{\mathrm{pred}}$, at the mutation sites $i$, grouped by the pLDDT of the predicted mutant structures. The error bars indicate the 95\% confidence interval estimated from $10^3$ bootstrap resamples.
(b) Pearson correlation coefficient between $\log \widetilde{D}_i$, and $\log \widetilde{D}_i^{\mathrm{pred}}$, at the mutation site $i$ for all predicted structures (red) and predicted structures with pLDDT $\geq 90$ (teal), calculated over residues satisfying $Z_i \geq Z_{\mathrm{cutoff}}$. The error bars indicate the 95\% confidence interval estimated from $10^3$ bootstrap resamples.
}
\label{fig:app5}
\end{figure}

In this appendix, we investigate whether the accuracy of AlphaFold3 in predicting mutation-induced structural deformation depends on its prediction confidence, as quantified by pLDDT. Figure~\ref{fig:app5}a shows the Pearson correlation coefficient between the normalized mutation-induced deformation, $\log \widetilde{D}_i$, and the normalized predicted mutation-induced deformation, $\log \widetilde{D}_i^{\mathrm{pred}}$, at the mutation sites $i$ for structures binned by pLDDT. The correlation remains approximately constant across pLDDT bins, indicating that higher pLDDT does not correspond to improved prediction of mutation-induced deformation. We further compared the prediction accuracy for all structures with that obtained after restricting the analysis to high-confidence structures with pLDDT $\geq 90$. As shown in Fig.~\ref{fig:app5}b, the correlation between $\log\widetilde{D}_i$ and $\log\widetilde{D}_i^{\mathrm{pred}}$ decreases with increasing $Z_{\mathrm{cutoff}}$ for both datasets, demonstrating that AlphaFold3 becomes less accurate for mutations that induce structural changes increasingly larger than the intrinsic structural fluctuations. Notably, restricting the analysis to structures with pLDDT $\geq 90$ does not improve this behavior; instead, the correlation decreases more rapidly than that obtained using all predicted structures. Thus, high pLDDT does not indicate increased accuracy in predicting mutation-induced structural deformation and does not account for the reduced predictive performance observed for strongly deforming mutations.

\section{Robustness of AlphaFold3 Prediction Accuracy Across Evaluation Metrics}
\label{app:f}

\begin{figure*}[!htbp]
\centering
\includegraphics[width=1\linewidth]{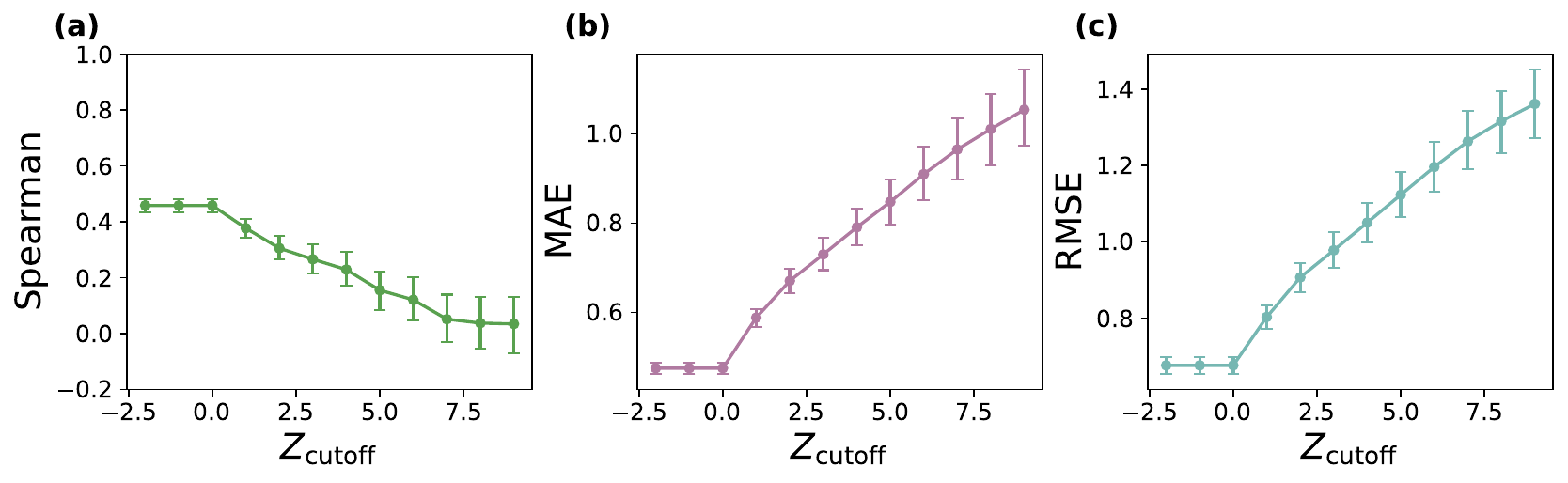}
\caption{
(a) Spearman correlation coefficient, (b) mean absolute error (MAE), and (c) root-mean-squared error (RMSE) between the normalized mutation-induced deformation, $\log \widetilde{D}_i$, and the normalized predicted mutation-induced deformation, $\log \widetilde{D}_i^{\mathrm{pred}}$, at the mutation site $i$, calculated for residues satisfying $Z_i \geq Z_{\mathrm{cutoff}}$. The error bars indicate the 95\% confidence interval estimated from $10^3$ bootstrap resamples.
}
\label{fig:app6}
\end{figure*}

In this appendix, we test whether the reduced prediction accuracy for strongly deforming mutations depends on the choice of the evaluation metric. We repeated the analyses using the Spearman correlation coefficient, mean absolute error (MAE), and root-mean-squared error (RMSE). As shown in Fig.~\ref{fig:app6}a, the Spearman correlation exhibits the same qualitative behavior as the Pearson correlation reported in the main text, decreasing as $Z_{\mathrm{cutoff}}$ increases. Consistently, both the MAE and RMSE increase with increasing $Z_{\mathrm{cutoff}}$ as shown in Fig.~\ref{fig:app6}b and c, indicating larger discrepancies between $\log \widetilde{D}_i$ and $\log \widetilde{D}_i^{\mathrm{pred}}$, for mutations that induce structural changes increasingly larger than the intrinsic structural fluctuations. Thus, the decrease in the AlphaFold3 prediction accuracy for strongly deforming mutations is robust to the choice of the evaluation metric.

\section{Boltz-2 yields similar prediction accuracy to AlphaFold3}
\label{app:g}

\begin{figure}[!htbp]
\centering
\includegraphics[width=1\linewidth]{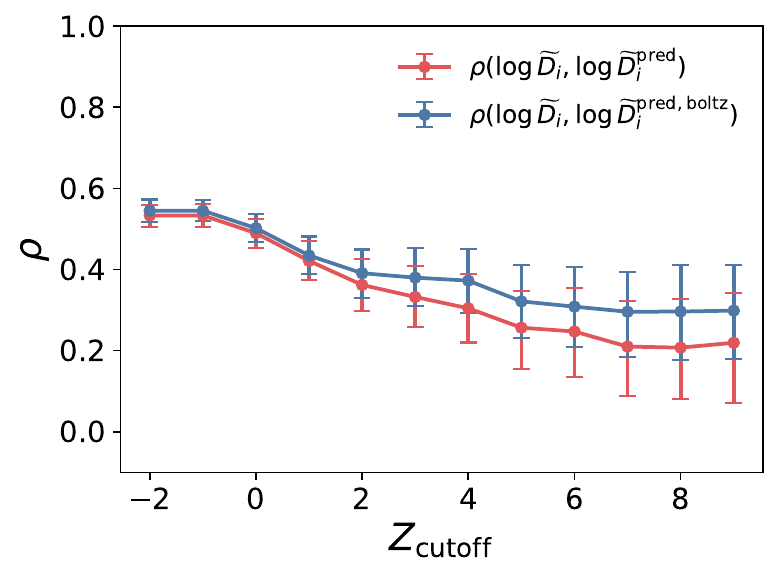}
\caption{
Pearson correlation coefficient between the normalized mutation-induced deformation, $\log \widetilde{D}_i$, and the predicted normalized mutation-induced deformation using AlphaFold3, $\log \widetilde{D}_i^{\mathrm{pred}}$ (red), and Boltz-2, $\log \widetilde{D}_i^{\mathrm{pred,boltz}}$ (blue), computed over mutation sites with $Z_i \geq Z_{\mathrm{cutoff}}$, plotted as a function of $Z_{\mathrm{cutoff}}$. The error bars indicate the 95\% confidence intervals estimated from $10^3$ bootstrap resamples.
}
\label{fig:app7}
\end{figure}

To assess whether our findings for AlphaFold3 generalize to other protein structure prediction models, we performed the analysis using structures predicted by Boltz-2. Fig.~\ref{fig:app7} compares the Pearson correlation coefficients between the experimentally measured normalized mutation-induced deformation, $\log \widetilde{D}_i$, and the corresponding predictions from AlphaFold3 $\log \widetilde{D}_i^{\mathrm{pred}}$ and Boltz-2 $\log \widetilde{D}_i^{\mathrm{pred,boltz}}$ plotted as a function of $Z_{\mathrm{cutoff}}$. The two protein structure prediction models exhibit similar correlations across the range of $Z_{\mathrm{cutoff}}$, with both showing limited accuracy in predicting normalized mutation-induced structural deformation, particularly for strongly perturbative point mutations. These results suggest that the observed limitations are not specific to AlphaFold3.

\newpage

\bibliography{references}

\end{document}